\documentclass[a4paper,11pt]{article}

\usepackage{pos}
\usepackage{bbold}
\usepackage{graphicx}
\usepackage{physics}
\usepackage{subcaption}
\usepackage{cleveref}
\usepackage{amsmath}
\usepackage{soul}

\newcommand{\beq}{\begin{equation}}
\newcommand{\eeq}{\end{equation}}
\newcommand{\beqs}{\begin{eqnarray}}
\newcommand{\eeqs}{\end{eqnarray}}
\newcommand{\nn}{\nonumber}

\title{\boldmath Spectroscopy of chimera baryons in a $Sp(4)$ lattice gauge theory}
\ShortTitle{Spectroscopy of chimera baryons in a $Sp(4)$ lattice gauge theory}

\author*[a]{Ho Hsiao}
\author[b]{Ed Bennett}
\author[c]{Deog Ki Hong}
\author[c,d]{Jong-Wan Lee}
\author[a,e]{C.-J. David Lin}
\author[b,f]{Biagio Lucini}
\author[g]{Maurizio Piai}
\author[h]{Davide Vadacchino}

\affiliation[a]{Institute of Physics, National Yang Ming Chiao Tung University, 1001 Ta-Hsueh Road, Hsinchu 30010, Taiwan}
\affiliation[b]{Swansea Academy of Advanced Computing, Swansea University (Bay Campus), Fabian Way, SA1 8EN Swansea, Wales, United Kingdom}
\affiliation[c]{Department of Physics, Pusan National University, Busan 46241, Korea}
\affiliation[d]{Institute for Extreme Physics, Pusan National University, Busan 46241, Korea}
\affiliation[e]{Center for High Energy Physics, Chung-Yuan Christian University, Chung-Li 32023, Taiwan}
\affiliation[f]{Department of Mathematics, Faculty of Science and Engineering, Swansea University (Bay Campus), Fabian Way, SA1 8EN Swansea, Wales, United Kingdom}
\affiliation[g]{Department of Physics, Faculty of Science and Engineering, 
Swansea University (Park Campus), Singleton Park, SA2 8PP Swansea, Wales, United Kingdom}
\affiliation[h]{Centre for Mathematical Science, University of Plymouth, Plymouth, PL4 8AA, United Kingdom}

\emailAdd{thepaulxiao.sc06@nycu.edu.tw}
\emailAdd{e.j.bennett@swansea.ac.uk}
\emailAdd{dkhong@pusan.ac.kr}
\emailAdd{jwlee823@pusan.ac.kr}
\emailAdd{dlin@nycu.edu.tw}
\emailAdd{b.lucini@swansea.ac.uk}
\emailAdd{m.piai@swansea.ac.uk}
\emailAdd{davide.vadacchino@plymouth.ac.uk}

\abstract{Chimera baryons are an important element of strongly coupled theories that provide a microscopic origin for UV complete composite Higgs models (CHMs), since they play the role of top partners in top partial compositeness. In a particular interesting realisation of CHMs based upon an underlying $Sp(4)$ gauge theory, such exotic objects are composed of two fermion constituents transforming on the fundamental, and one on the 2-index antisymmetric representations. We perform lattice computations of the chimera baryon spectrum in the quenched approximation. We present preliminary results for the masses of various chimera baryons with different quantum numbers, including the one interpreted as the top partner. We test the  technology needed for future calculations with dynamical fermions.}

\FullConference{%
The 39th International Symposium on Lattice Field Theory,\\
8th-13th August, 2022,\\
Rheinische Friedrich-Wilhelms-Universität Bonn, Bonn, Germany
}

\begin{document}
\maketitle
	
\section{Introduction}\label{sec:intro}

Our interest in the study of $Sp(4)$ gauge theories arises from the contexts of CHMs~\cite{KAPLAN1984183,Georgi:1984af,Dugan:1984hq}. References~\cite{Ferretti:2013kya,Barnard:2013zea} show that  the $Sp(4)$ gauge theory with two fermionic fields in fundamental and three in 2-index antisymmetric representations is a CHM candidate with $SU(4)/Sp(4)$ coset. It is the minimal model containing $5$ pseudo Nambu-Goldstone bosons; $4$ of them can be interpreted as the Higgs doublet in the Standard Model (SM) and $1$ can be made heavy. An exotic object, called chimera baryon, can mix with the standard model top quark, providing mass to the latter through partial compositeness. This mechanism for generating the top mass was firstly introduced in Ref.~\cite{Kaplan:1991dc}. The first lattice study of chimera baryons was presented in Ref.~\cite{Ayyar:2018zuk} for a $SU(4)$ gauge theory.

Our collaboration has been conducting studies of $Sp(2N)$ gauge theories by employing well-developed lattice technology~\cite{Bennett:2017kga,Lee:2018ztv,Bennett:2019jzz,Bennett:2019cxd,Bennett:2020hqd,Bennett:2020qtj,Bennett:2021mbw,Bennett:2022yfa,Bennett:2022ftz} (see Refs.~\cite{Bennett:2021mbw,Bennett:2022yfa} for the discussion and preliminary results for chimera baryons). We aim to launch the mass spectrum calculations of $Sp(4)$ gauge theory with dynamical Dirac fermions in two distinct representations. As a preliminary step before doing highly demanding dynamical lattice simulations, we study the spectrum of chimera baryons with quenched fermions to guide us on bare lattice parameter space when generating gauge configurations. In this article, we present the interpolating operator of chimera baryons and our preliminary results in the quenched approximation. 

\section{Interpolating operators and measurements}

In order to build a spin-1/2 composite fermion neutral under $Sp(4)$ gauge group, the simplest way is to combine two fundamental fermions (denoted by $Q$) and one 2-index antisymmetric fermion (denoted by $\Psi$). The interpolating operator reads
\beq\label{eq:ocb}
\mathcal{O}_{\rm CB}^{\gamma} (x) \equiv
\left (
{Q^{i\,a}}_\alpha(x)  \Gamma^{1\,\alpha\beta} {Q^{j\,b}}_\beta(x)
\right)
\Omega_{ad}\Omega_{bc} \Gamma^{2\, \delta\gamma} {\Psi^{k\,cd}}_{\gamma}(x)\,,
\eeq
where $a,\,b,\,c,\,d$ are colour indices, $i,\,j,\,k$ are flavor indices, $\alpha,\,\beta,\,\delta,\,\gamma$ are spinor indices, $\Gamma^{1,2}$ are combinations of gamma matrices and $\Omega$ is the $4\times4$ symplectic matrix. The $SU(3)_c$ SM gauge group is a subgroup of the flavour symmetry acting on $\Psi^k$. The Dirac conjugate of the interpolating operator is 
\beq
\overline{\mathcal{O}_{\rm CB}^{\gamma}} (x) \equiv \overline{\Psi^{k\,cd}}_\delta (x) \Omega_{cb}\Omega_{da}  \Gamma^{2\, \delta\gamma}
\left (
{\overline{Q^{j\,b}}_\beta}(x)  \Gamma^{1\,\beta\alpha} \overline{Q^{i\,a}}_\alpha(x)
\right)\,.
\eeq
Considering the case $i \neq j$, the zero momentum two-point correlation function is
\beqs
C^{ \gamma\gamma^\prime}(t) &\equiv& \sum_{\vec{x}} \langle \mathcal{O}_{\rm CB}^{\gamma}(x) \overline{\mathcal{O}_{\rm CB}^{\gamma^\prime}}(0) \rangle \nn \\
&=& - \sum_{\vec{x}} \left ( \Gamma^2 {S_{\Psi}^{k\,cd }}_{c^\prime d^\prime} (x,0)  \overline{\Gamma^{2}} \right)_{\gamma\gamma^\prime} \Omega_{cb}\Omega^{b^\prime c^\prime} \Omega_{ad}\Omega^{d^\prime a^\prime}
\nn\\
&& ~~~
 \times\Tr \left [ \Gamma^{1}  S^{b}_{Q\,\,\,b^{\prime}}(x,0) \overline{\Gamma^{1}} S_{Q\,\,\,\,a^\prime}^{a}(x,0) \right ],
\eeqs
where we define $x\equiv(t,\vec{x})$ and $\overline{\Gamma} \equiv \gamma^0 \Gamma^\dagger \gamma^0$, the trace is over the spinor indices, and the fermion propagators are defined as
\beq
S^{\,i\,a}_{Q\,\,\,\,b\,\alpha\beta}(x,y) = 
\langle Q^{i\,a}_{\,\,\,\,\,\alpha}(x) \overline{Q^{i\,b}}_{\beta}(y) \rangle
~{\textrm{and}}~S^{\,k\,ab}_{\Psi\,\,\,\,\,\,\,\,cd\,\alpha\beta}(x,y) =
 \langle \Psi^{k\,ab}_{\,\,\,\,\,\,\,\,\,\,\alpha}(x) 
\overline{\Psi^{k\,cd}}_{\beta}(y) \rangle\,.
\label{eq:fermion_prop}
\eeq
The interpolating operator of top partner chimera baryon has been written down in Ref.~\cite{Bennett:2022yfa}. One uses $(\Gamma^1,\,\Gamma^2)\equiv (C\gamma^5,\,\mathbb{1})$. Other choices of gamma matrices can be found in Ref.~\cite{Cossu:2019hse} for the $SU(4)$ gauge theory. In analogy with QCD, we denote as $\Lambda$ the state with quantum numbers $(J,R) = (1/2,5)$, where $J$ is the spin and $R$ denotes the irreducible representation of the flavour group acting on $Q^i$. This state provides a natural candidate as a top partner~\cite{Gripaios:2009pe}. At large Euclidean time, the $2$-point correlation function for this $\Lambda$-type operator asymptotically behaves as 
\beq\label{eq:ceo}
C(t) \xrightarrow{t\rightarrow \infty} P_e \left[ c_ee^{-m_et} + c_oe^{-m_o(T-t)} \right] - P_o \left[ c_oe^{-m_ot} + c_ee^{-m_e(T-t)} \right],
\eeq 
where $P_{e}$ and $P_{o}$ are parity projectors defined as
\beq
P_e \equiv \frac{1}{2}(1+\gamma^0)
~{\textrm{and}}~
P_o \equiv \frac{1}{2}(1-\gamma^0)\,.
\eeq
We denote as $c_e$ and $c_o$ the coefficients, $m_e$ and $m_o$ the mass of parity even and odd states, respectively. $T$ is the temporal lattice size.
As we see from Eq.~(\ref{eq:ceo}), the backward propagator of the even projected correlator, $C_{e}(t) \equiv P_{e} C(t)$, decays with the same mass as the forward propagator of the odd projected correlator, $C_{o}(t) \equiv P_{o} C(t)$, and vice versa. Thus, we make use of two projections to improve the statistics by averaging with the function 
\beqs\label{eq:ctilde}
\tilde{C}(t) = \frac{1}{2} \left [ C_{e}(t) + C_{o}(T-t) \right ]
\xrightarrow{t\rightarrow \infty} \tilde{c}_e e^{-m_et} + \tilde{c}_o e^{-m_o(T-t)}.
\eeqs

We also measure other types of chimera baryons, for completeness. Other possible low lying states are $\Sigma$: $ (J,R) = (1/2,10)$ and $\Sigma^*$: $ (J,R) = (3/2,10)$, which share the same interpolating operator constructed using $(\Gamma^1,\,\Gamma^2)\equiv (C\gamma^\mu,\,\mathbb{1})$ in Eq. (\ref{eq:ocb}). In general, the $\Sigma$ state could also play the role of the top partner~\cite{Banerjee:2022izw}. The two-point function is defined as
\beq
C_{ij}^{ \gamma\gamma^\prime} (t) \equiv \sum_{\vec{x}} \langle \mathcal{O}^{\gamma}_{\textrm{CB}\,,i}(x) \overline{\mathcal{O}^{\gamma^\prime}_{\textrm{CB}\,,j}}(0) \rangle \,,
\eeq
where $i$ ($j$) refers to the case of $\Gamma^1 = C\gamma^i$ with $i = 1,2,3$.
The spin-1/2 and -3/2 states are isolated by employing projectors (see Ref.~\cite{UKQCD:1996ssj} for the analogous treatment of heavy baryon in QCD):
\beq
P^{1/2}_{ij} \equiv \frac{1}{3} \gamma_i \gamma_j\,
~{\textrm{and}}~
P^{3/2}_{ij} \equiv \delta_{ij} -\frac{1}{3}\gamma_i \gamma_j\,,
\eeq
for spin-1/2 states $C_{1/2}^{ \gamma\gamma^\prime}(t) \equiv P^{1/2}_{ij}C^{ \gamma\gamma^\prime}_{ji}(t)$, and for spin-3/2 states $C_{3/2}^{ \gamma\gamma^\prime}(t) \equiv P^{3/2}_{ij}C_{ji}^{ \gamma\gamma^\prime}(t)$.


\section{Mass spectrum}

Ensembles are generated using the same setup of an earlier study of the mesonic spectroscopy with quenched fermions~\cite{Bennett:2019cxd}. We aim at exploring how chimera baryon masses depend on the fermion masses. To improve the signal, we employ Wuppertal (Gaussian) smearing~\cite{Gusken:1989ad} and APE smearing~\cite{Falcioni:1984ei} to construct operators with better overlap with the ground state. 
\subsection{Parity and spin projections}

\begin{figure}[t!]
\centering
     \begin{subfigure}[b]{0.49\textwidth}
         \centering
            \includegraphics[width=\textwidth]{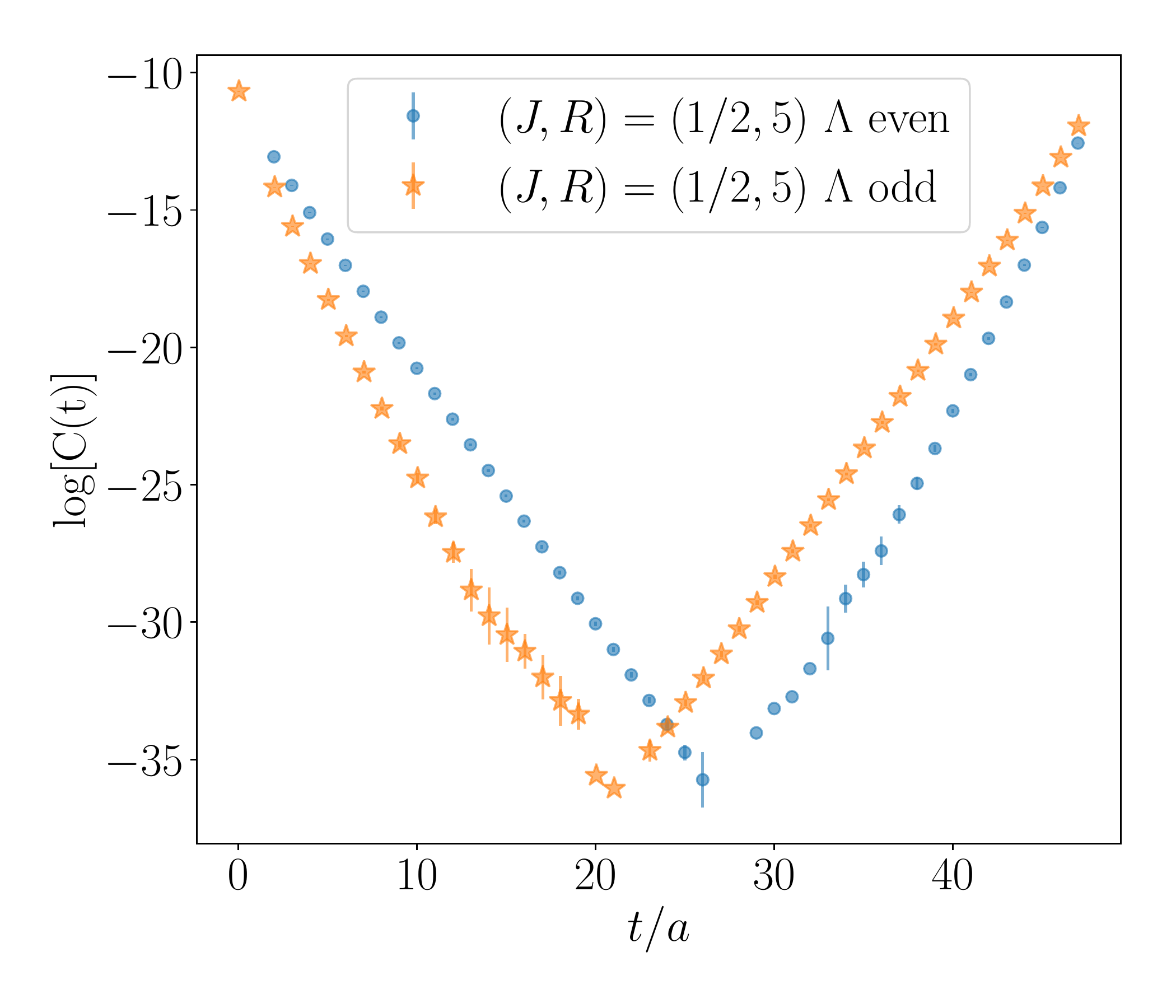}
            \caption{} \label{fig:MCB_eolnc}
     \end{subfigure}
     \hfill
     \begin{subfigure}[b]{0.49\textwidth}
         \centering
         \includegraphics[width=\textwidth]{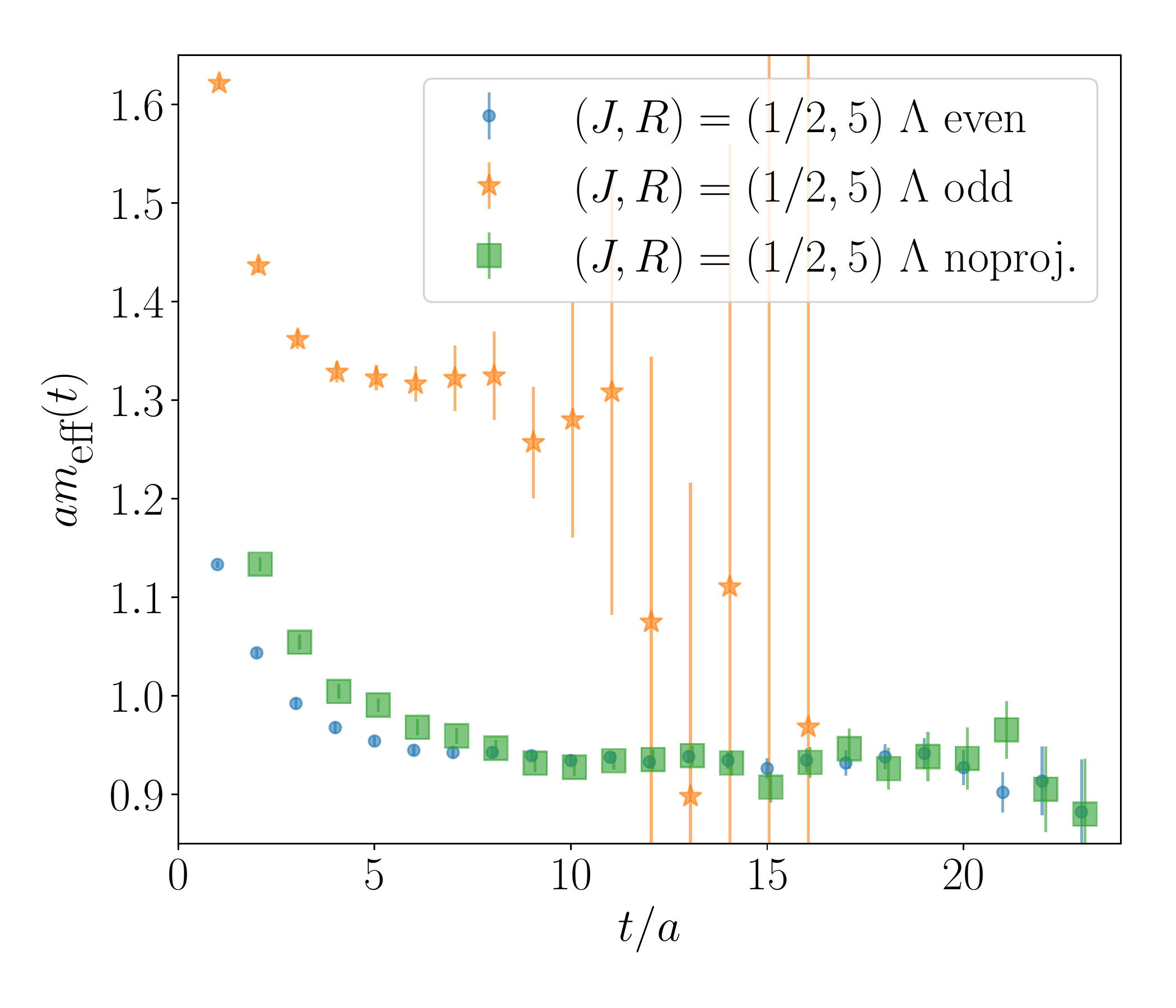}
         \caption{} \label{fig:MCB_eo}
       \end{subfigure}
    \caption{{\bf(a)} Log plot of correlators and {\bf(b)} effective mass plot comparison with parity projections in units of lattice spacing $a$. The blue circles and orange stars denote the even and odd projected correlators, respectively. The green squares in the right panel denote the original correlators without projection. The calculations are performed on a $48\times 24^3$ lattice (QB1 in Ref.~\cite{Bennett:2019cxd}). }
\end{figure}

\begin{figure}[t!]
	\centering
	\includegraphics[width=0.49\textwidth]{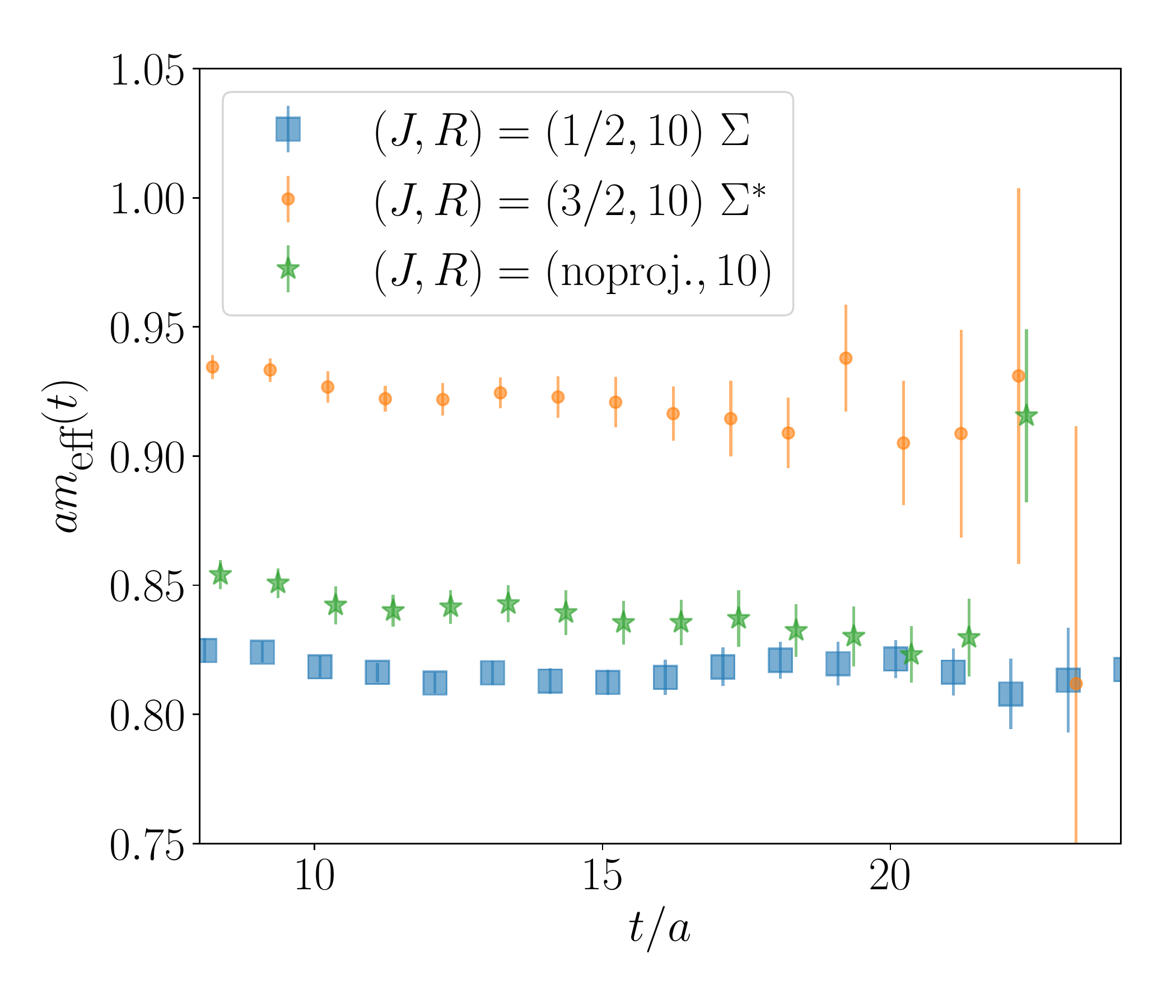}
	\caption{Comparison of effective mass plots between two spin states and the state without spin projection. At large Euclidean time, the spin-1/2 state (blue squares), which is compatible with the mass obtained without spin projection (green stars), is lighter than spin-3/2 state (orange points).}
	\label{fig:MCB_spin}
\end{figure}

We demonstrate the conjugate propagation properties of even and odd states in Fig.~\ref{fig:MCB_eolnc}. The correlators of two parity states behaves as Eq.~(\ref{eq:ceo}). For instance, the forward propagating state with even projection is the same as the backward propagating state with odd projection. Because of such an asymmetry in the correlation function, we do not use the usual cosh function for the extraction of masses. In principle, we could extract the masses for both parity even and odd states by fitting data to Eq.~(\ref{eq:ctilde}). However, for this preliminary study we focus on the lightest one, and the masses are determined by performing a single exponential fit to $\tilde{C}(t)$ over the time range of the plateau in the effective mass defined as
\beq
m_{\text{eff}}(t) \equiv \ln\left( \frac{\tilde{C}(t)}{\tilde{C}(t+1)} \right ).
\eeq
As shown in Fig.~\ref{fig:MCB_eo}, we find that the mass of the parity-even state is statistically the same as the mass without projection at large Euclidean time. Since the mass of the parity-even state is lower than that of the parity-odd state and measured with a higher precision, we focus on the former. We further find that parity projection significantly reduces statistical errors of the parity-even masses. Similar phenomena can be found with the spin projections, see Fig.~\ref{fig:MCB_spin}. The effective mass at large Euclidean time without spin projection is compatible with that of the spin-1/2 state but has larger error, and the spin-3/2 state is heavier than the spin-1/2 state.

\subsection{Mass hierarchy of chimera baryons and fermion mass dependence}

\begin{figure}
\centering
     \begin{subfigure}[b]{0.49\textwidth}
         \centering
            \includegraphics[width=\textwidth]{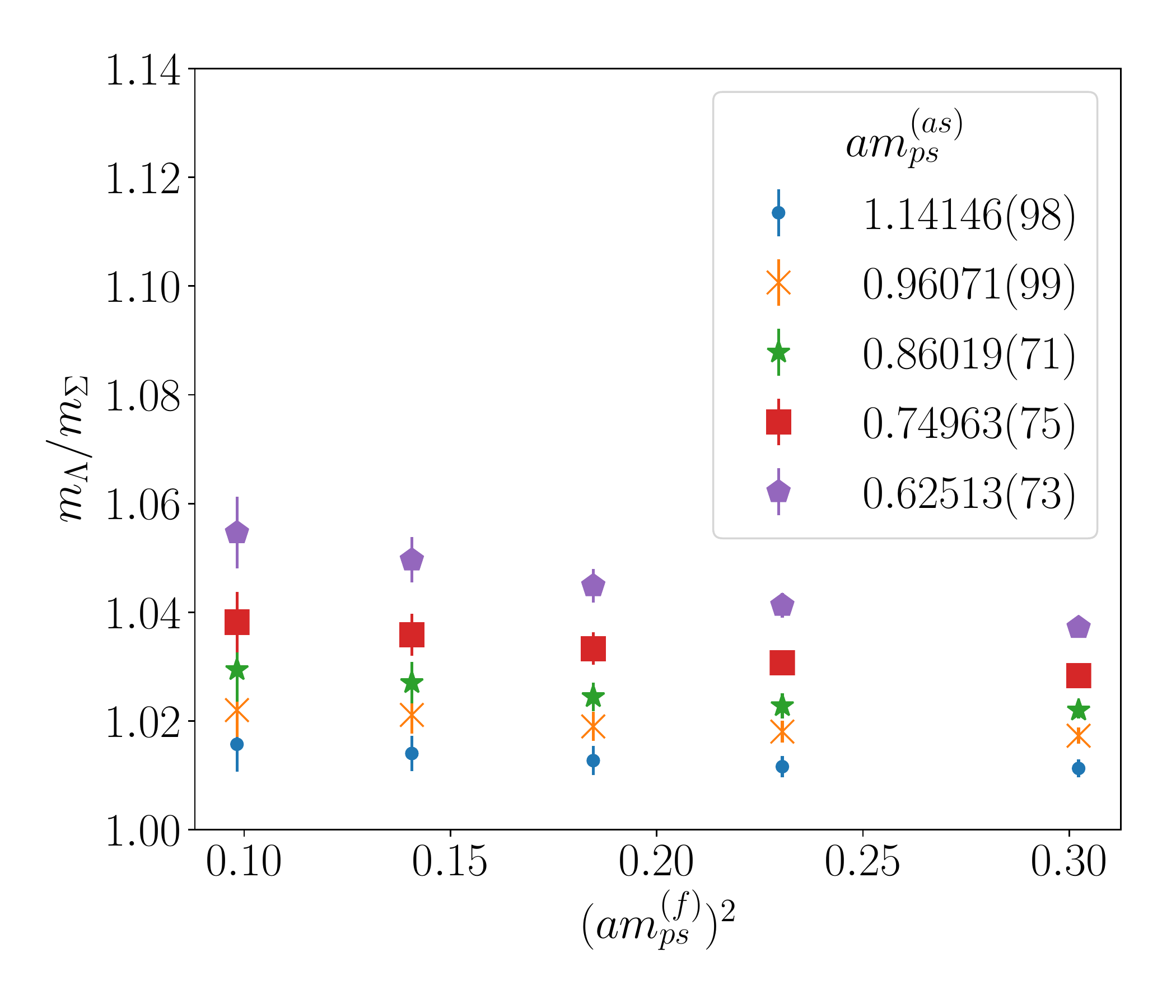}
            \caption{} \label{fig:MCB_rls}
     \end{subfigure}
     \hfill
     \begin{subfigure}[b]{0.49\textwidth}
         \centering
         \includegraphics[width=\textwidth]{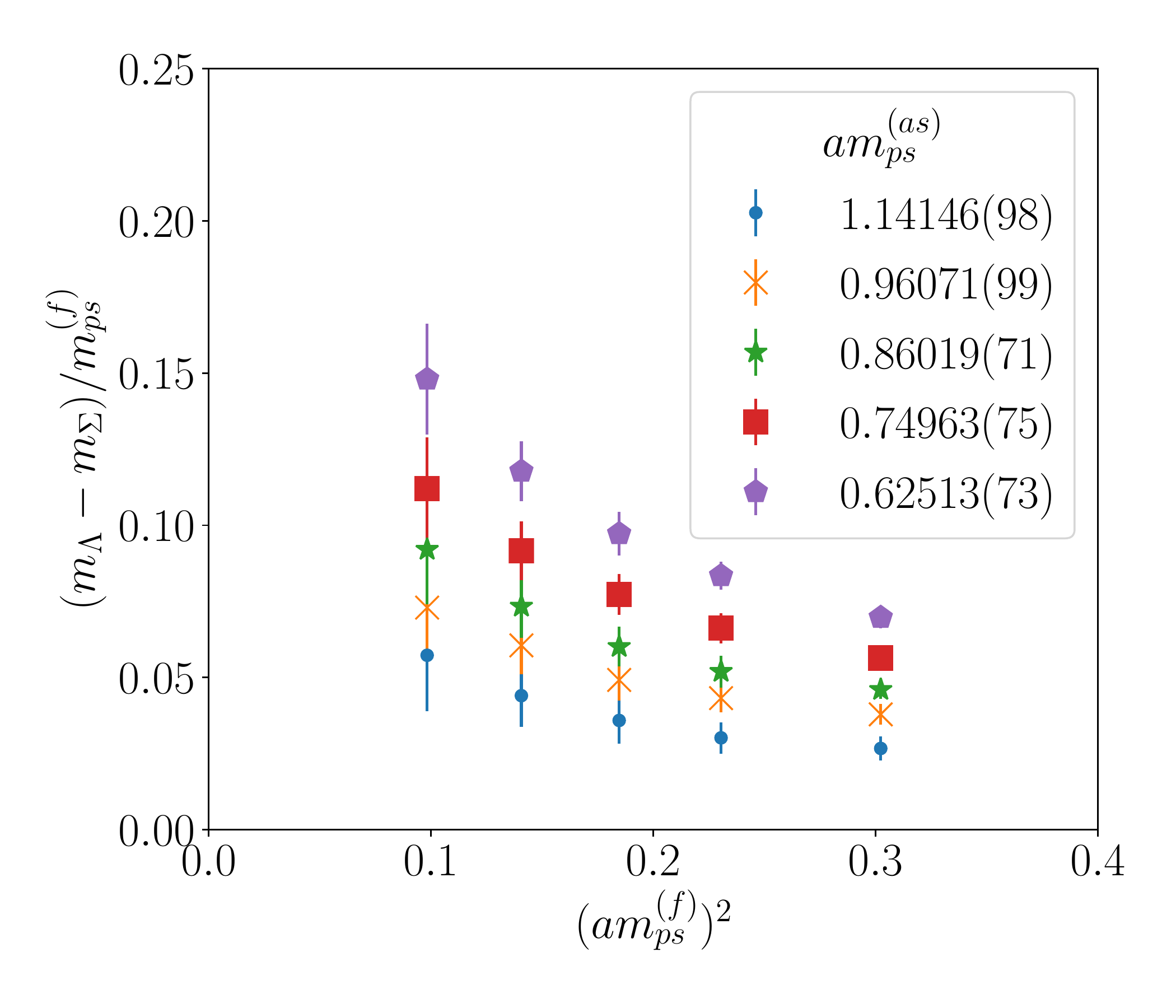}
         \caption{} \label{fig:MCB_rlsps}
       \end{subfigure}
    \caption{Plot of {\bf(a)} the mass ratio $m_{\Lambda}/m_{\Sigma}$ and {\bf(b)} the ratio $(m_{\Lambda}-m_{\Sigma})/m^{\rm (f)}_{\rm ps}$ against fundamental pseudoscalar meson mass square $(am^{\rm (f)}_{\rm ps})^2$. Different colours and shapes represent measurements at different antisymmetric pseudoscalar meson masses.}\label{fig:MCB_Rs}
\end{figure}

In order to study how the constituent fermion masses affect the chimera baryon masses, we examine their dependence on the pseudoscalar meson mass. We extract the chimera baryon masses by fitting the correlators with a single exponential function, while using the cosh function for the analysis of meson correlators. As a first step, we vary the bare masses on one single ensemble (the smallest lattice volume with largest lattice spacing, QB1 in Ref.~\cite{Bennett:2019cxd}). One interesting feature we immediately discovered is that the $\Lambda$ chimera baryon is not the lightest state; as demonstrated by Fig.~\ref{fig:MCB_rls}. Under the assumption that $\Lambda$ is the top partner, it should be a stable state, for model-building purposes. We examine the naive possible decay process of $\Lambda$ into $\Sigma$ and a pseudoscalar meson by checking the ratio $(m_{\Lambda}-m_{\Sigma})/m^{\rm (f)}_{\rm ps}$. As long as the ratio is less than one, the $\Lambda$ chimera baryon would be stable.  As shown in Fig~\ref{fig:MCB_rlsps}, this is the case for all our current simulations.
\begin{figure}
\centering
     \begin{subfigure}[b]{0.49\textwidth}
         \centering
            \includegraphics[width=\textwidth]{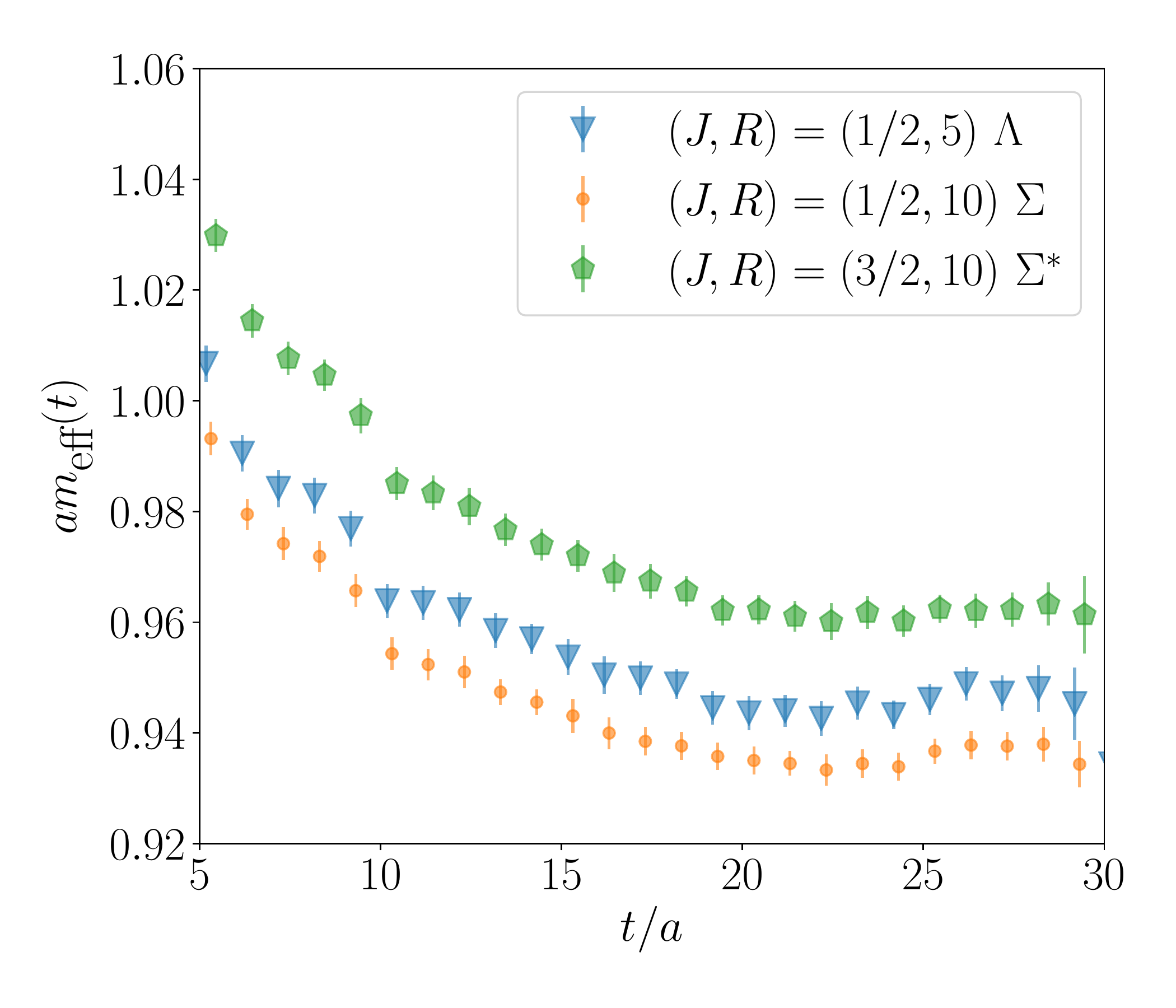}
            \caption{} \label{fig:MCB_mh1}
     \end{subfigure}
     \hfill
     \begin{subfigure}[b]{0.49\textwidth}
         \centering
         \includegraphics[width=\textwidth]{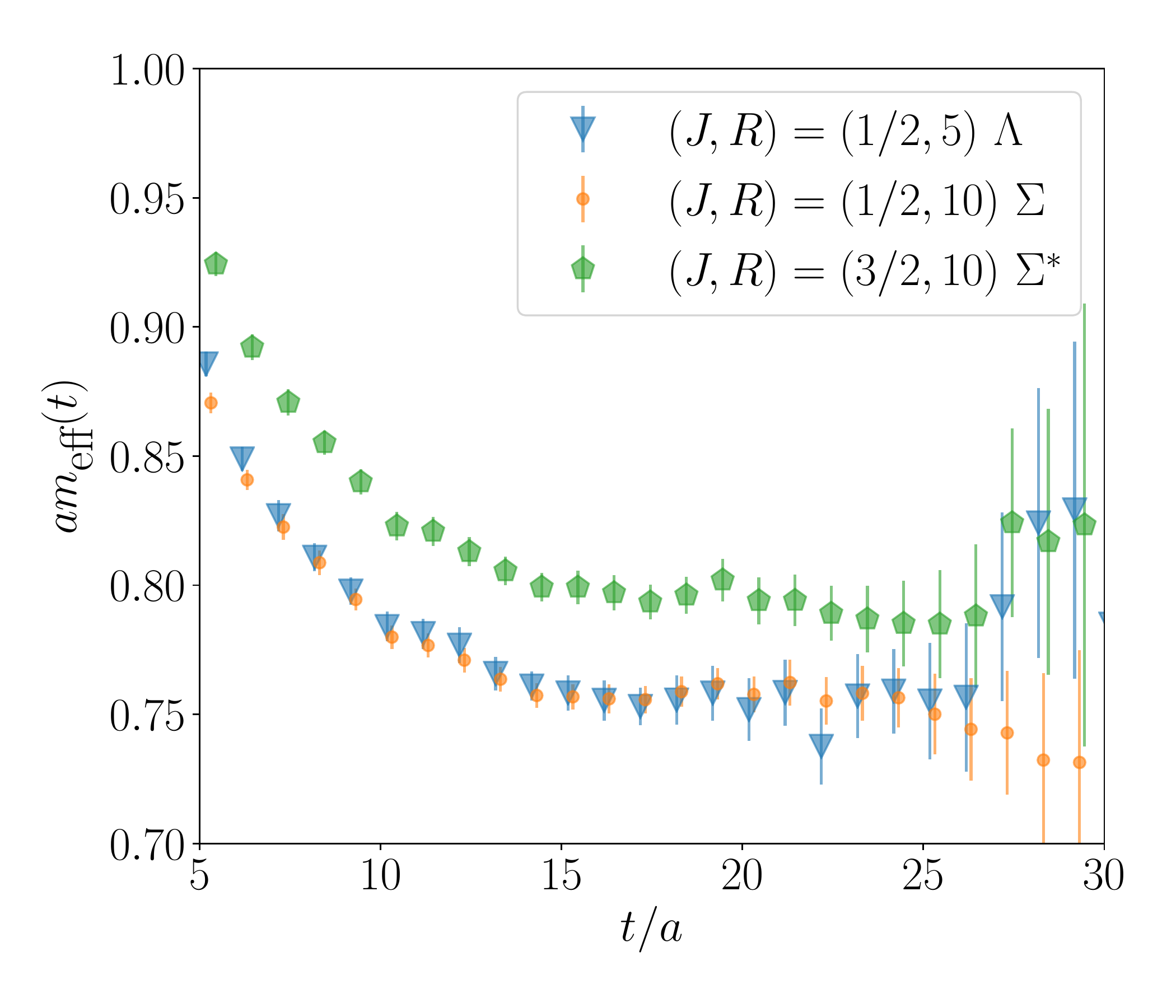}
         \caption{} \label{fig:MCB_mh4}
       \end{subfigure}
    \caption{Effective mass plot of chimera baryons calculated with different fundamental bare masses {\bf(a)} $m_0^{(f)} = -0.6$ and {\bf(b)} $m_0^{(f)} = -0.69$, at fixed antisymmetric bare mass $m_0^{(as)} = -0.81$, on a $60\times48^3$ finer lattice (QB4 in Ref.~\cite{Bennett:2019cxd}). 
    }\label{fig:MCB_mh}
\end{figure}

Figure \ref{fig:MCB_rls} also shows that $m_{\Lambda}/m_{\Sigma}$ becomes smaller when the fundamental pseudoscalar meson mass increases. Similar effects are observed when increasing the antisymmetric pseudoscalar meson mass. It should be noted that the heaviest pseudoscalar meson mass in the antisymmetric representation is already above the cut-off scale. We further illustrate such effects on our finest lattice with a volume, $60\times48^{3}$ (QB4 in Ref. [7]).   Figure~\ref{fig:MCB_mh} exhibits that the $\Sigma$ baryon is clearly lighter than the $\Lambda$ baryon at $m^{\rm (as)}_{\rm ps}/m^{\rm (f)}_{\rm ps} \sim 1.8$. On the other hand, if we decrease the fundamental bare mass while holding fixed the antisymmetric one, the masses of $\Lambda$ and $\Sigma$ become compatible with one another when $m^{\rm (as)}_{\rm ps}/m^{\rm (f)}_{\rm ps} \sim 4.46$. This suggests that at sufficient large antisymmetric pseudoscalar meson mass and relative small fundamental pseudoscalar meson mass, there is possibility that the $\Lambda$ chimera baryon is the lightest state. We are currently completing the investigation on all the ensembles. Our study of the relation between pesudoscalar meson masses and baryon mass hierarchy has implications for constructing viable CHMs with partial compositeness.

\begin{figure}

\centering
     \begin{subfigure}[b]{0.49\textwidth}
         \centering
            \includegraphics[width=\textwidth]{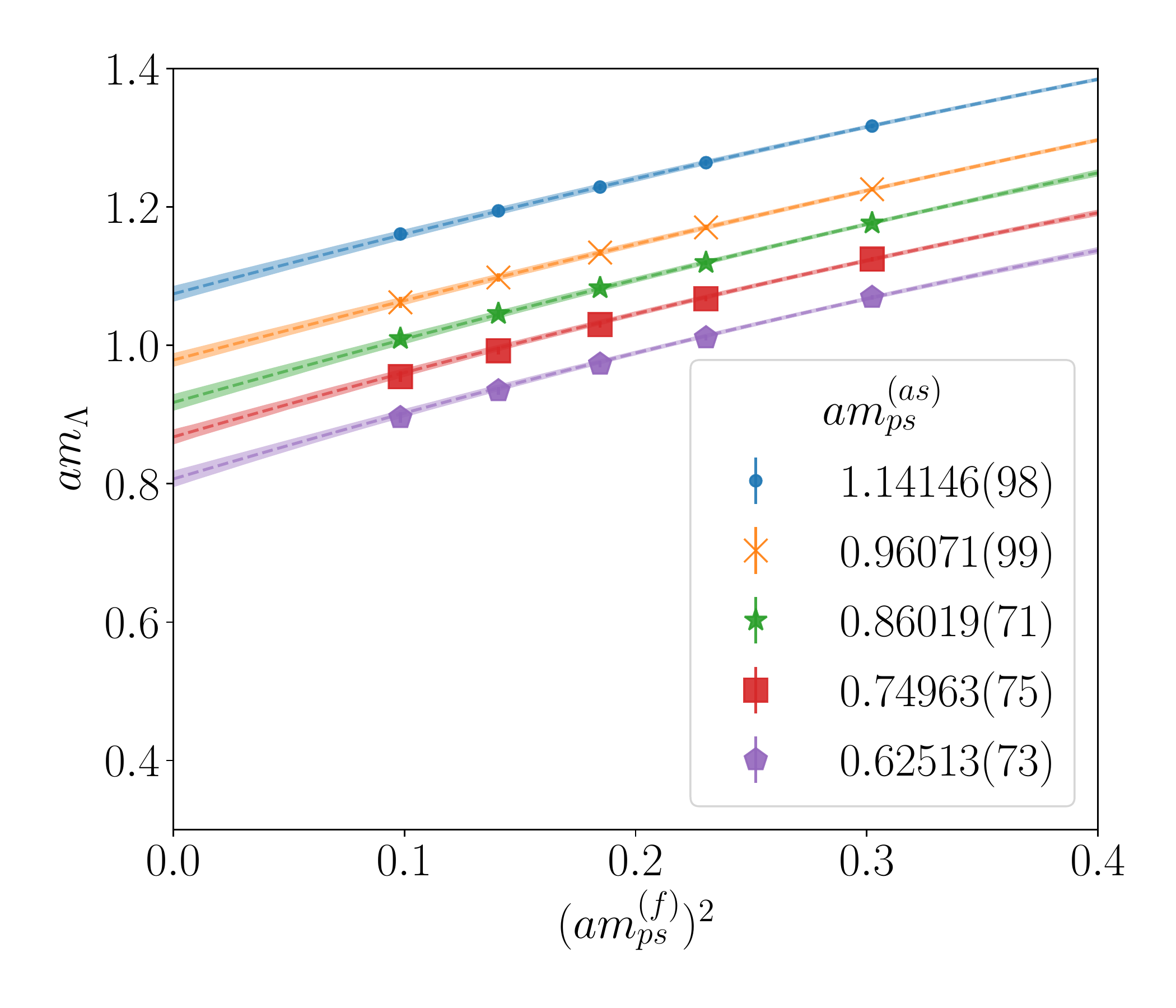}
            \caption{} 
     \end{subfigure}
     \hfill
     \begin{subfigure}[b]{0.49\textwidth}
         \centering
         \includegraphics[width=\textwidth]{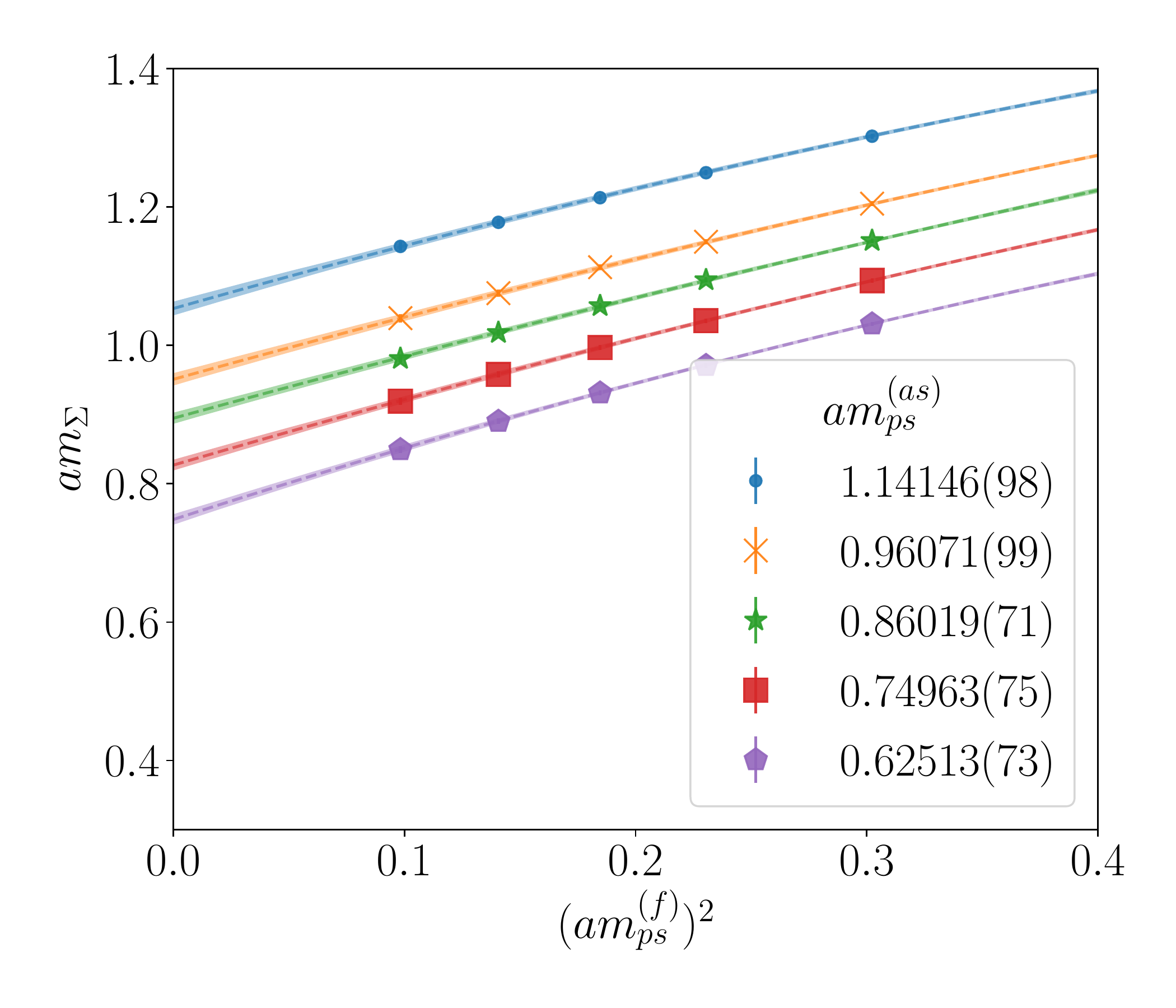}
         \caption{} 
       \end{subfigure}
       \hfill
     \begin{subfigure}[b]{0.49\textwidth}
         \centering
         \includegraphics[width=\textwidth]{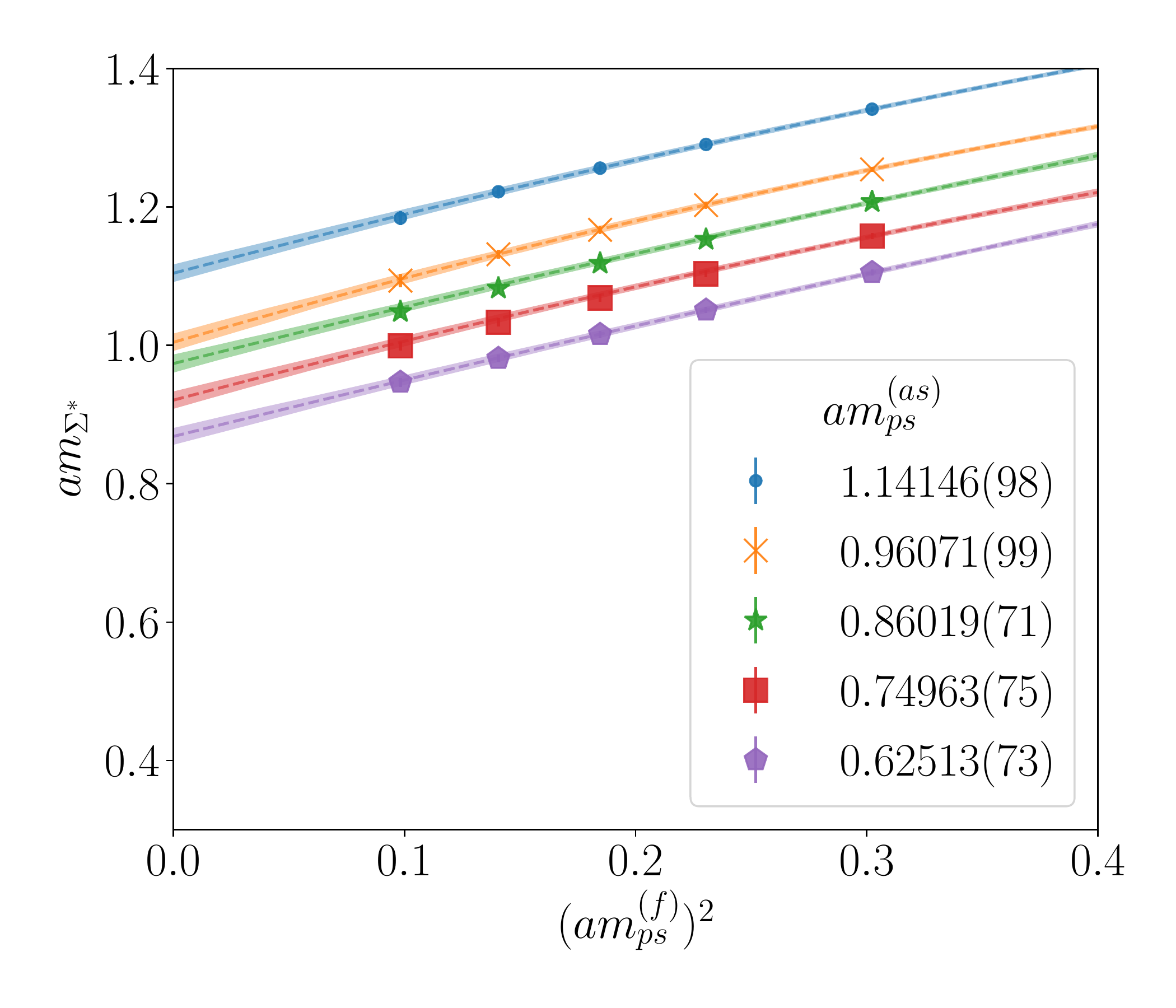}
         \caption{} 
       \end{subfigure}
       \hfill
       \caption{Mass extrapolation of {\bf(a)} $\Lambda$, {\bf(b)} $\Sigma$ and {\bf(c)} $\Sigma^*$ chimera baryons towards massless fundamental pesudoscalar meson at fixed antisymmetric pesudoscalar meson masses (different colours and shapes), which are performed by fitting a polynomial function with respect to $(am^{\rm (f)}_{\rm ps})^2$.}\label{fig:mcb_expt}
\end{figure}

We now turn our attention to the mass dependence of an individual chimera baryon. We perform an extrapolation towards the limit of massless fundamental pseudoscalar meson. As a first attempt with one single ensemble (QB1), the extrapolation of the chimera baryon mass is performed at fixed antisymmetric pesudoscalar meson mass with a naive polynomial function including $(am^{\rm (f)}_{\rm ps})^2$ and $(am^{\rm (f)}_{\rm ps})^4$ terms. See Fig.~\ref{fig:mcb_expt} for the results. In future work, we will make this exploratory study more systematic and rigorous by using baryonic chiral perturbation theory~\cite{Jenkins:1990jv}.

\section{Conclusion and outlook}

In the $Sp(4)$ gauge theory with two fundamental and three antisymmetric Dirac fermions, a candidate to UV-complete CHMs with partial compositeness, chimera baryons play an important role as the top partners. In this contribution, we have presented the interpolating operators for chimera baryons, and the analysis of relevant correlators for extracting their masses. We have reported preliminary results on the masses of three chimera baryon states, $\Lambda$: $(J,R) = (1/2,5)$, $\Sigma$: $(J,R) = (1/2,10)$ and $\Sigma^*$: $(J,R) = (3/2,10)$. One important result for model building we found is that the mass hierarchy of chimera baryons depends on the pseudoscalar meson masses. These are preliminary results, mostly measured on one quenched ensemble. We are currently performing analysis on all the quenched ensembles available to us, that will serve as a guide to future investigations with fully dynamical simulations.

\acknowledgments
We thank Gabriele Ferretti for useful comments.
The work of H.~H. and C.~J.~D.~L. is supported by the Taiwanese MoST Grant No. 109-2112-M-009 -006 -MY3.
The work of E. B. has been funded in part by the UKRI Science and Technology Facilities Council (STFC) Research Software Engineering Fellowship EP/V052489/1. 
The work of D.~K.~H. was supported by Basic Science Research Program through the National Research Foundation of Korea (NRF) funded by the Ministry of Education (NRF-2017R1D1A1B06033701).
The work of J.~W.~L is supported by the National Research Foundation of Korea (NRF) grant funded by the Korea government (MSIT) (NRF-2018R1C1B3001379).
The work of B.~L. and M.~P. has been supported in part by the STFC Consolidated Grants No. ST/P00055X/1 and No. ST/T000813/1. B.~L. and M.~P. received funding from the European Research Council (ERC) under the European Union’s Horizon 2020 research and innovation program under Grant Agreement No. 813942.
The work of B.~L. is further supported in part by the Royal Society Wolfson
Research Merit Award No. WM170010 and by the Leverhulme Trust Research Fellowship No. RF-2020-4619. 
The work of D.~V. is supported in part the Simons Foundation under the program “Targeted Grants to Institutes” awarded to the Hamilton Mathematics Institute.
Numerical simulations have been performed on the Swansea SUNBIRD
cluster (part of the Supercomputing Wales project) and AccelerateAI A100 GPU system,
on the local HPC clusters in Pusan National
University (PNU) and in National Yang Ming Chiao Tung University
(NYCU), and on the DiRAC Data Intensive service at Leicester. 
The Swansea SUNBIRD system and AccelerateAI are part funded
by the European Regional Development Fund (ERDF) via
Welsh Government. The DiRAC Data Intensive service at Leicester is operated 
by the University of Leicester IT Services, which forms part of the STFC DiRAC HPC Facility (www.dirac.ac.uk). The DiRAC Data Intensive service equipment at Leicester was funded by BEIS capital funding via STFC capital
grants ST/K000373/1 and ST/R002363/1 and STFC DiRAC Operations grant ST/R001014/1. 
DiRAC is part of the National e-Infrastructure.

{\bf Open Access Statement-} For the purpose of open access, the author has applied a Creative Commons Attribution (CC BY) licence to any Author Accepted Manuscript version arising. Raw data and the analysis workflow used will be made available concurrently with the forthcoming journal publication of this work.

\end{document}